\documentclass[conference]{IEEEtran}
\IEEEoverridecommandlockouts
\usepackage{cite}
\usepackage{amsmath,amsthm,amssymb,amsfonts}
\usepackage{algorithmic}
\usepackage{graphicx}
\usepackage{textcomp}
\usepackage{xcolor}

\def\BibTeX{{\rm B\kern-.05em{\sc i\kern-.025em b}\kern-.08em
    T\kern-.1667em\lower.7ex\hbox{E}\kern-.125emX}}
\begin{document}

\title{Update Rate, Accuracy, and Age of Information in a Wireless Sensor Network\\
}

\author{\IEEEauthorblockN{Xinlu Dai, Cyril Leung}
\IEEEauthorblockA{\textit{Department of Electrical and Computer Engineering} \\
\textit{The University of British Columbia}\\
Vancouver, Canada \\
Emails: xinludai@ece.ubc.ca, cleung@ece.ubc.ca}
}

\maketitle

\begin{abstract}
Age of Information (AoI), namely the time that has elapsed since the most recently delivered packet was generated, 
is receiving increasing attention with the emergence of many real-time applications that rely on the exchange of time-sensitive information. AoI captures the freshness of the information from the perspective of the destination.
The term ``accuracy of information" is used to assess how close the estimate at the destination is to the parameter value measured by the sensor. In this paper, the mean square error (MSE) is used to evaluate the accuracy of information.
We focus on a single sensor that monitors a time-sensitive physical process, which is modelled as a random walk.
Whenever the state of the random walk changes by more than a specified threshold, the sensor generates a status update packet and transmits it to the destination. When no update packet is received, the destination assumes that the state of the process has not changed.
We study the problem of finding the minimum update rate under AoI and accuracy of information constraints. More specifically, we derive analytical expressions for the update rate, the AoI, and the MSE. 
\end{abstract}


\section{Introduction}
In the past few years, there has been a significant increase in the number of applications that rely on the exchange of time-sensitive information for monitoring and control. Examples include autonomous vehicular systems, smart factories, environmental monitoring, sensor networks, etc. For many applications, it is crucial to keep the information as fresh as possible because its value decreases rapidly with time. 

Much work has been done to improve throughput, reduce packet delay and mitigate the effects of transmission errors.
The packet delay has been commonly used to study time-sensitive information. 
However, it does not fully capture the freshness of status updates at the destination. For example, it does not deal with the question of when updates should be generated at the source.

To address this shortcoming, the Age of Information (AoI) has been proposed to quantify the freshness of information at the destination \cite{kaul_1}. It is defined as the time that has elapsed since the generation of the most recently delivered packet at the destination \cite{kaul_1}. 

Stochastic models for the arrivals of status updates at the source are commonly used for studying the AoI \cite{kaul_1,maice,kadota_sto}. 
Three types of update packet generation are commonly adopted in the AoI literature \cite{kaul_1,maice,update_wait,lazy,kadota_1,kadota_throu,kadota_sto,xingran}: (1)~Stochastic generation of status updates at the source according to some random process, e.g., a Poisson process. (2)~Deterministic generation of update packets at a fixed rate. (3)~A new update packet is generated as needed. This last update generation model is assumed in this paper.


We will use the term ``accuracy of information" to describe how close the value estimated at the destination is to the parameter value generated by the source.
To evaluate the accuracy of information, we adopt the average mean squared error (MSE).

Different applications have different performance requirements. For example, systems involving human safety generally require stricter AoI targets than outdoor environmental monitoring; banking systems need higher accuracy than smart home systems. The update rate is closely related to energy consumption. Generally, a lower update rate results in a reduced energy budget. Energy savings are especially important in applications involving battery-powered devices located in remote areas. However, a lower update rate may adversely affect AoI. Therefore, there is a need to balance different performance metrics such as AoI, update rate, and accuracy of information. 

The trade-off between AoI and other performance metrics has been previously studied.
In \cite{lazy,update_wait,maice}, the minimization of the AoI with respect to the update rate is discussed.
In~\cite{kadota_throu,through_1}, the optimization of AoI with throughput constraints is explored. In~\cite{through_2,through_3}, the trade-off between AoI and throughput is examined.
AoI with energy harvesting sources is studied in \cite{lazy,eh_1,eh_2,eh_3}. 

However, the trade-offs between update rate, AoI, and accuracy of information have not been studied much. Ideally, a low AoI, a high accuracy, and a low update rate are desired. In this paper, we study the relationships among these 3 performance measures. The results can be used to determine the minimum update rate needed to achieve specified AoI and accuracy of information requirements. 

\section{System Model}
Consider a communication link with one source\nobreakdash-destination pair in a single\nobreakdash-hop wireless network. 
A sensor monitors an underlying time\nobreakdash-varying process and intends to share information (sensor data) about the state of the process with the destination. 
Time is assumed to be slotted with a slot duration normalized to unity, with slot index $n \in \{0, 1,2,\cdots, N\}$. The system model is now summarized.

\begin{itemize}
    \item \textbf{Sensor data:}
    The sensor data change according to a 1\nobreakdash-D random walk on the integers \cite{biology,gnp,finance}, with process state changes occurring at the beginning of every time slot. 
 The change, $X_i$, in the walker's position (state of the random walk) at time $i, i=0,1,2,\ldots$ is a random variable (r.v.) with the following distribution 
\begin{align}
X_{i}=\left\{\begin{array}{ll}
+1 & \text {with probability } p  \\
-1 & \text {with probability } q \\
0 & \text {with probability } 1-p-q
\end{array}\right.
\end{align}
where $p,q \in (0,1)$, $p+q \in (0,1)$, and 
$\{X_i\}$ is a sequence of independent, identically distributed (iid) r.v.s.
Denote the state (sensor output) at time $n$ by $S_n$, where $S_0$ is the initial state. Then we have 
\begin{equation}
    S_n = S_0+ \sum_{i=1}^{n} X_{i}, \; n = 1, 2, \ldots
\end{equation}
        
  
    \item{\textbf{Packet transmission time:}
    The transmission and propagation time of a status update packet is assumed to be one time slot, with no queuing delay.}
    
    \item {\textbf{Update scheme:}
    Two relative boundaries, one at $-T$ and the other at $T$ relative to the position (state) $S$ reported in the most recent update, where $T$ is a positive integer, are used by the sensor to adjust the update rate. For simplicity, we will refer to $T$ as the (relative) boundary threshold. If the sensor sends an update packet when the walker is at position (state) $S$, it will only generate a new update packet when the walker reaches state $S-T$ or $S+T$. 
    When an update packet is generated (at the beginning of time slot $n$), it is immediately sent to the destination.
    
    Let $u(n)$ be an indicator function that is equal to 1 if the walker reaches the relative boundary at $-T$ or at $T$ at the beginning of time slot $n$; otherwise $u(n)=0$. When $u(n)=1$, the sensor generates a new packet and transmits it to the destination. At the beginning of time slot 0, the sensor sends an update packet containing the initial walker position to the destination.
    
    \item {\textbf{Channel noise:}
    For simplicity, we assume that the channel noise is negligible, i.e.,  update packets are delivered error-free to the destination.
    }}
    \end{itemize}
\section{Performance Evaluation}
\label{sec:performance}

\subsection{Update Rate}
\label{subsec:ur}
The update rate is the frequency at which the sensor generates update packets. For example, if the sensor generates an update packet every 4 time slots, the update rate is 0.25 packet\,/\,time slot.

An update cycle refers to the period starting from the delivery time of an update packet to the delivery time of the next update packet; delivery times occur at the end of time slots.
We denote the random variable for the length of (i.e. the number of time slots in) an update cycle by $L$, where $L \geq 1$.

The average update rate, $\lambda$, is the inverse of the average update cycle length. Unless otherwise noted, the average update rate will be simply referred to as the update rate.

Since the state evolution follows a random walk, the length of each update cycle is equal to the time starting from state $S_n$ to absorption at state $S_n-T$ or $S_n+T$. 
Since a random walk is a Markov process, the next state only depends on the current state and is independent of the history. Once the boundary is reached at the beginning of time slot $n$, the current state $S_n$ becomes the initial state of the random walk for the next update cycle. Therefore, the average length of the cycle is equal to the expected time starting from the origin to the absorbing time at state $T$ or $-T$.

We now describe how to derive this expected time. At the beginning of each time slot, the state increases by $1$ with probability $p$, decreases by $1$ with probability $q$ and remains unchanged with probability $(1-p-q)$.


Let $D_z$ denote the expected time to absorption when the random walk starts at state $z, 0<z<d$, and ends when either state~$0$ or state~$d$ is reached.
Then,
\begin{equation}
\begin{aligned}
D_{z} & =p (D_{z+1}+1)+q (D_{z-1}+1)+(1-p-q)(D_z +1)\\
& = p D_{z+1}+q D_{z-1}+(1-p-q) D_z + 1, \quad 0<z<d.
\end{aligned}
\end{equation}
Therefore, 
\begin{equation}
    D_{z+1} -\frac{p+q}{p} D_z + \frac{q}{p} D_{z-1} = -\frac{1}{p}, \label{nonhomo}
\end{equation}
with the boundary conditions 
\begin{equation}
D_{0}=0, \quad D_{d}=0. \label{boundary}
\end{equation}
Equation~\eqref{nonhomo} is a second-order difference equation with constant coefficients. 

It can be shown that 
\begin{equation}
    \mathbb E \left[ L \right] = D_T=\left\{\begin{array}{ll}
\displaystyle \frac{T(p^T-q^T)}{(p-q)(p^T+q^T)}, & \text {when } p \neq q  \vspace{1ex} \\
\displaystyle \frac{T^2}{2p}, & \text{when } p=q
\end{array}\right.
\label{EL}
\end{equation}

Details of the derivation are provided in \cite{Dai_2022}. The update rate is then given by
\begin{equation}
    \lambda  = \frac{1}{\mathbb E \left[L\right]} \\
    \label{update rate}
\end{equation}

Equation~\eqref{update rate} shows that $\lambda$ is a decreasing function of $T$ as to be expected.

%
%
\subsection{Age of Information (AoI)}
\label{subsec:AoI}
In a slotted system, AoI is defined as the number of time slots that have elapsed since the generation time of the most recent status update packet received at the destination \cite{kadota_slot}.
The AoI at the beginning of time slot $n$ is denoted by $h(n)$.
The packet generation takes place only at the beginning of a time slot and the transmission time is one slot. As a result, if there is no transmission during time slot $n$, i.e., the update indicator $u(n)=0$, then $h(n+1)$ increases by one, i.e., $h(n+1)=h(n)+1$. In contrast, if there is a transmission during slot $n$, i.e., $u(n)=1$, then $h(n+1)$ decreases to one because the packet is one time slot old after the transmission. 

We will refer to the time-average AoI simply as the average AoI. Due to the randomness of the sensor data, there exists randomness in the update times. Therefore, the expected normalized sum AoI $(\mathrm{ENSAoI})$ per time slot is considered. The ENSAoI from slot 0 to slot $N$ is defined as the expectation of the ratio of the sum of AoI over $N+1$ slots to the number of time slots \cite{kadota_throu,xingran}, i.e.,
\begin{equation}
    \text{ENSAoI}= \mathbb E \left[ \frac{1}{N+1} \sum_{n=0}^{N} H(n) \right],
    \label{ENSAoI_defi}
\end{equation}
where $H(n)$ is the random variable representing the AoI for time slot $n$.
Specifically, we discuss the infinite-horizon normalized sum AoI (NSAoI), i.e.,
\begin{equation}
    \text{NSAoI}=  \lim_{N \to \infty} \frac{1}{N+1} \sum_{n=0}^{N} H(n).
    \label{NSAoI_inf}
\end{equation}

\begin{figure}[!t]
\centering
\includegraphics[scale=0.38]{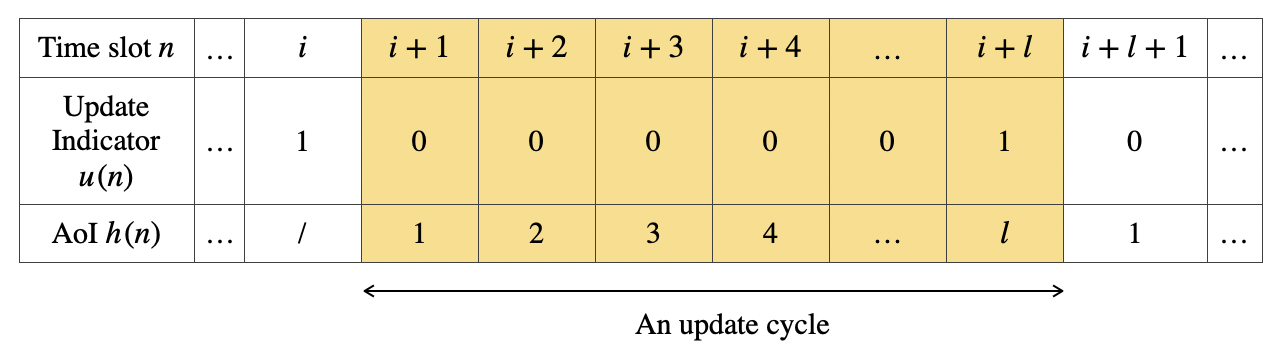}
\caption{An example of AoI evolution in an update cycle of length $l$.}
\label{fig:aoi-cycle}
\end{figure}

The evolution of $h(n)$ in an update cycle of length $l$ is shown in Fig.~\ref{fig:aoi-cycle}. We note that the sequence of AoI values $h(n)$ is an arithmetic progression starting from 1 to $l$, with a common difference of 1. Thus, the sum AoI for an update cycle of length $l$ is given by
\begin{equation}
    \begin{aligned}
        \text{AoI}_\text{cycle sum} \triangleq Z(l) = {1+2+3+\cdots+l} = \frac{l(l+1)}{2}.
    \end{aligned}
\end{equation}

Consider a sequence of $K$ update cycles. The normalized sum AoI for the $K$ update cycles is given by
\begin{footnotesize}
\begin{equation}
    \text{NSAoI}(K) = \frac{\text{sum AoI for the $K$ update cycles}}
    {\text{sum of the number of time slots in the $K$ update cycles}}.
\end{equation}
\end{footnotesize}
The infinite-horizon NSAoI is obtained by setting $K \to \infty$, i.e.,
\begin{equation}
    \text{NSAoI} = \lim_{K \to \infty} \text{NSAoI}(K).
\end{equation}

Let the different cycle lengths, arranged in increasing order, in the $K$ cycles be denoted by $l_1, l_2,\cdots$, where $l_1<l_2<\cdots$. Let $C_1,C_2,\cdots$ be the random variables representing the number of cycles of length $l_1, l_2, \cdots$, respectively. We note that 
\begin{equation}
    K=C_1+C_2  + \cdots,
\end{equation}
where $C_1, C_2,\cdots$ are independent.
Denote the probability distribution of $L$ by $P_L(l)$. 
Using the probability distribution of the time to absorption in \cite{cox}, we can write
\begin{footnotesize}
\begin{align}
   &P_L(l)= p_{-T}(l)+p_{T}(l) \notag \\ 
    &=\left[\left(\frac{q}{p}\right)^{\frac{T}{2}}+ \left(\frac{q}{p} \right)^{-\frac{T}{2}} \right] \frac{\sqrt{p q} }{T} \sum_{\nu=1}^{2T-1} \frac{(-1)^{\nu+1} \sin \left(\frac{\nu \pi}{2}\right) \sin \left(\frac{\nu \pi}{2T}\right)}{(s_{\nu})^{l-1}},
    \label{pl}
\end{align}
\end{footnotesize}
where
\begin{small}
\begin{equation}
    s_{\nu}=\frac{1}{1-p-q+2 \sqrt{p q}\cos\left(\frac{\nu \pi}{2T}\right)},\nu=1,2, \ldots, 2T-1.
    \label{sv_T}
\end{equation}
\end{small}

Since the frequency of an update cycle of length $l$ is equal to $P_L(l)$ as $K \to \infty$, we have
\begin{equation}
    \lim_{K \to \infty}\frac{C_m}{K} = P_L(l_m), \; m=1,2,\cdots.
\end{equation}
Then,
\begin{small}
\begin{subequations}
\begin{align}
    & \text{NSAoI} = \lim_{K \to \infty} \frac{ C_1 Z(l_1)+ C_2 Z(l_2)+ \cdots + C_m Z(l_m)+\cdots}
    {C_1 l_1 + C_2 l_2 + \cdots + C_m l_m+\cdots} \\
    &= \frac{P_L(l_1)  Z(l_1)+
    P_L(l_2)  Z(l_2)+ \cdots +
    P_L(l_m)  Z(l_m)+\cdots}
    {P_L(l_1) l_1 +P_L(l_2) l_2 + \cdots +P_L(l_m) l_m+\cdots} \\
    &= \frac{\sum_{l=1}^{+\infty}P_{L}(l)\cdot  \frac{l (l+1)}{2}}{\sum_{l=1}^{+\infty} P_{L}(l) \cdot l}\\
    &= \frac{1}{2} \left(1+\frac{\mathbb E[L^2]}{\mathbb E[L]} \right).
    \label{NSAoI}
\end{align}
\end{subequations}
\end{small}

\subsection{Accuracy of Information}
\begin{figure}[!t]
\centering
\includegraphics[scale=0.42]{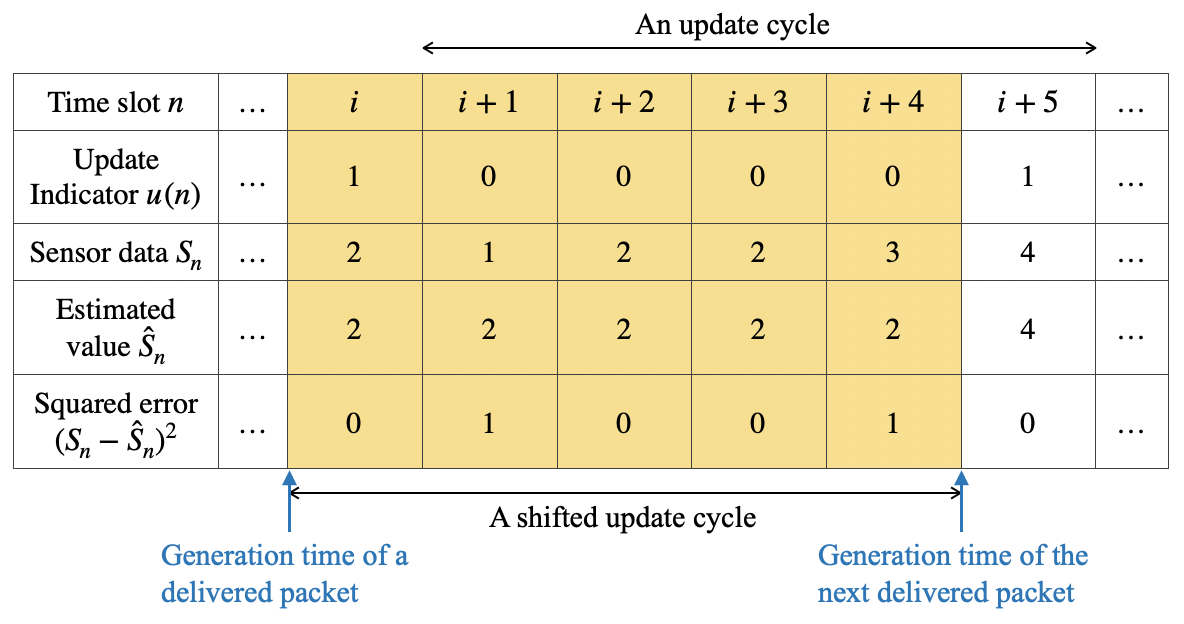}
\caption{An illustration of a shifted update cycle and estimation error with $T=2$, and $p+q \neq 1$.}
\label{fig:mse-cycle}
\end{figure}
In this section, we assess the accuracy of information using the estimation error. 
For an arbitrary time slot $n$, let $S_n$ denote the actual value monitored by the sensor and $\widehat{S}_n$ denote the estimated value at the destination. 
The destination cannot wait until it receives a future update before estimating the current state, as it needs to operate with minimal delay. We assume the use of the following simple estimator
\begin{equation}
    \widehat{S}_n = \left\{\begin{array}{cl}
\widehat{S}_{n-1}, &  \text { if } u(n)=0 \\
S_n, &  \text { if } u(n)=1
\end{array}\right.
\label{S-evolution}
\end{equation}
where $\widehat{S}_0 = S_0$.

As illustrated in Fig.~\ref{fig:mse-cycle}, the estimated values $\widehat{S}_n$ in a shifted update cycle, which starts from the generation time of a delivered packet to the generation time of the next delivered packet, are identical.
For ease of calculation, we use a shifted update cycle. The length of the shifted update cycle is the same as that of the update cycle and is denoted by $L$ as well.

The estimation error for an arbitrary slot $n$ is defined as
\begin{equation}
    \text{Error}_n = |S_n -\widehat{S}_n|,
\end{equation}
and the MSE from slot $0$ to slot $N$ as
\begin{equation}
    \text{MSE}_N = \frac{1}{N+1} \sum_{n=0}^{N} \left( S_n - \widehat{S}_{n} \right)^2.
\end{equation}
Due to the randomness in sensor data, we calculate the expected long-term MSE (EMSE) as 
\begin{equation}
    \text{EMSE}= \mathbb E \left[\lim_{N \to \infty} \frac{1}{N+1} \sum_{n=0}^{N} \left( S_n - \widehat{S}_{n} \right)^2\right].
\end{equation}
Unlike the AoI in an update cycle which forms an arithmetic sequence, the pattern of errors in a shifted update cycle is not evident. 

Let the first state of an arbitrary shifted update cycle be $S_m$ at time $m$. This cycle will end right before the state first reaches $S_{m}+T$ or $S_{m}-T$. 
Given the Markov property of the random walk, $S_{m}$ can be assumed to be 0 when calculating the squared error. Then the shifted cycle ends right before the state first hits $T$ or $-T$. This process repeats for the next cycle. 
Therefore, for the purpose of calculating the estimation error, the random walk we use is equivalent to a {\em modified} random walk in which the walker starts from 0 and every time the boundary is hit, the current state is reset to 0. We refer to this modified random walk as $\{S_n'\}$.
The range of states in every shifted update cycle is $\left[-T+1, T-1\right]$. Since the first state in a shifted update cycle, $S_m'=0$, is sent to the destination and there is no other update in this cycle, the estimated states in this cycle are the same as $\widehat{S_m'}$ and $\widehat{S_m'}=S_m'=0$ (see \eqref{S-evolution}). Thus the estimation error of an arbitrary slot $n$ in this shifted cycle is given by $|S_n'-\widehat{S_n'}|=|S_n'|$, i.e.
\begin{equation}
    \text{EMSE}= \mathbb E \left[\lim_{N \to \infty}\frac{1}{N+1} \sum_{n=0}^{N} \left( S_n'\right)^2\right].
    \label{emse}
\end{equation}

We are interested in determining 
the steady-state distribution of $S_n'$.
The EMSE steady-state (or stationary) distribution can be expressed as
\begin{equation}
    \text{EMSE} = \mathbb E \left[\lim_{n \to \infty} \left( S_n'\right)^2 \right].
    \label{emse-long}
\end{equation}

For the {\em modified} random walk, let the current state be denoted by $i$, $i \in \mathcal{I}$, where $\mathcal{I}=\{-T+1,-T+2,\cdots,T-2,T-1\}$. 
The probability of moving from state $i$ to state $j$, is denoted by
\begin{equation}
    p_{i,j} = \mathbb P (S'_{n+1}=j \mid S'_n =i), 
     i,j \in \mathcal{I}, n = 0,1,2,\cdots
\end{equation}
If $-T+1<i<T-1$, the next possible state is $i+1$, $i-1$ or $i$ itself. 
If $i=-T+1$, the next state could be $i+1$, 0 or $i$ but not $i-1=-T$ since moving left in this case corresponds to the event of returning to state 0. Similarly, if $i=T-1$, the next state could be 0, $i-1$ or $i$.

The transition probabilities can be expressed as a $(2T-1) \times (2T-1)$ matrix $P$ in which the $(i,j)$th element is $p_{i,j}$. Then,
\begin{tiny}
\begin{equation}
\setlength{\arraycolsep}{1pt}
\begin{aligned}
P &= \left[ {\begin{array}{ccccccccccc}
    1-p-q   &   p   &   0   &  \cdots & q& \cdots&   0&  0&  0\\
    q   &   1-p-q   &   p   &  \cdots & 0 & \cdots&   0&  0&  0\\
    0   &   q   &   1-p-q   &  \cdots & 0 & \cdots&  0&  0&  0\\
    \vdots &  \vdots & \vdots & \ddots & \vdots & \ddots &  \vdots & \vdots & \vdots\\
    0 & 0 & 0 &  \cdots & 0 & \cdots&    1-p-q&  p&  0\\
    0 & 0 & 0 &  \cdots & 0  & \cdots&   q&  1-p-q&  p\\
    0 & 0 & 0 &  \cdots & p & \cdots&  0&  q&  1-p-q\\
  \end{array} } \right]
\end{aligned}
\end{equation}
\end{tiny}

This Markov Chain is irreducible as all states communicate. 
For each state $i \in \mathcal{I}$, 
let
\begin{equation}
    \pi_i^{(n)} = \mathbb P (S'_n = i), \quad n=0,1,2,\cdots,
    \label{pi-defi}
\end{equation}
be the probability that the Markov Chain is in state $i$ at time $n$. 
We can represent the state probability distribution of $S'_n$ as a $1 \times (2T-1)$ row vector
\begin{equation}
    \pi^{(n)} = \left(\pi_{-T+1}^{(n)},\; \pi_{-T+2}^{(n)},\;\cdots,\;\pi_{T-2}^{(n)},\; \pi_{T-1}^{(n)}\right).
    \label{pi-vector}
\end{equation}
Then,
\begin{equation}
    \pi^{(n)}=\pi^{(n-1)}P=\pi^{(n-2)}P^2=\cdots=\pi^{(0)}P^{n},
    \label{pi_n}
\end{equation}
where $\pi^{(0)}$ is the initial probability distribution of $S'_0$. 

If a finite Markov Chain is irreducible, equation
\(
    \pi=\pi P 
\)
has a unique solution which is the stationary distribution of the Markov Chain \cite{cox}.

Note that in the special case when $p+q=1$ and $T$ is even, the Markov chain is periodic, i.e. the chain can only move from a state $S$
to state $S \pm 2k$, $k=0,1,2,\cdots$ after an even number of steps whereas it can only move from a state $S$ to state $S \pm (2k-1)$ after an odd number of steps. 

Let the stationary distribution for $n$ odd (even) be denoted by $\pi_{odd}$ ($\pi_{even})$. 
 Assuming that the probability of $n$ being even or odd is the same, we have
\begin{equation}
    \pi = \left\{\begin{array}{cl}
\pi_{odd}, &  \text { with probability } \frac{1}{2} \vspace{1ex}\\
\pi_{even}, &  \text { with probability } \frac{1}{2}
\end{array}\right.
\label{pi-special}
\end{equation}
For all other cases, there is one identical stationary distribution no matter whether $n$ is odd or even.
From \eqref{emse-long},~\eqref{pi-defi},~\eqref{pi_n},~\eqref{pi-special}, for $T \geq 2$, the EMSE can be calculated as
\begin{subequations}
\begin{align}
    &\text{EMSE} = \sum_{i \in \mathcal{I}} i^2 \mathbb P (S'_\infty = i)\\
    &= \left\{\begin{array}{cl}
\displaystyle \frac{1}{2}\sum_{i \in \mathcal{I}} i^2 \left(\pi_{i,odd}+\pi_{i,even}\right),\!&p\!+\!q\!=\!1 \text { and } T \text{ is even }\vspace{1ex}\\
\displaystyle \sum_{i \in \mathcal{I}} i^2 \pi_i, &  \text { otherwise}
\end{array}\right.\label{emse-cal}
\end{align}
\end{subequations}
\section{Numerical Results}
\subsection{Update rate}
\begin{figure}[!t]
\centering
\includegraphics[scale=0.2]{ 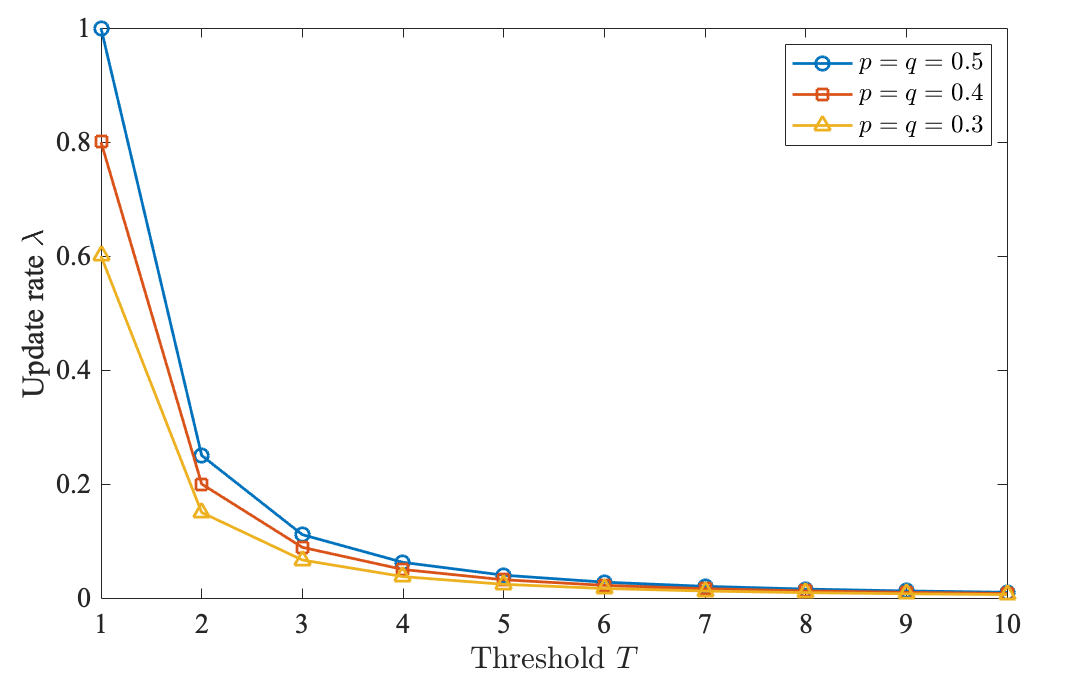}
\caption{Update rate $\lambda$ as a function of the threshold $T$,  with $p=q=0.5$, 0.4, 0.3.}
\label{fig:ur1}
\end{figure}
The update rate can be evaluated using ~\eqref{update rate}.
Fig.~\ref{fig:ur1} shows the update rate $\lambda$ as a function of the threshold $T$ for $p=q$, i.e., a symmetric random walk. As $T$ increases from 1 to 10, $\lambda$ first decreases quite rapidly for $T \leq 4$. The decrease is more gradual for $T \geq 5$. The decreasing trend is to be expected, as a larger $T$ makes the boundaries more difficult to be reached, resulting in a longer average update cycle. For a fixed $T$, a higher value of $1-p-q$ results in a lower update rate because the state is more likely to remain the same, making it more difficult to reach the boundaries. For $p\neq q$, the update rate as a function of $T$ curve is generally similar to that for $p=q$.

\subsection{AoI}
\label{subsec:numerical-aoi}

\begin{figure}[!t]
\centering
    \includegraphics[scale=0.2]{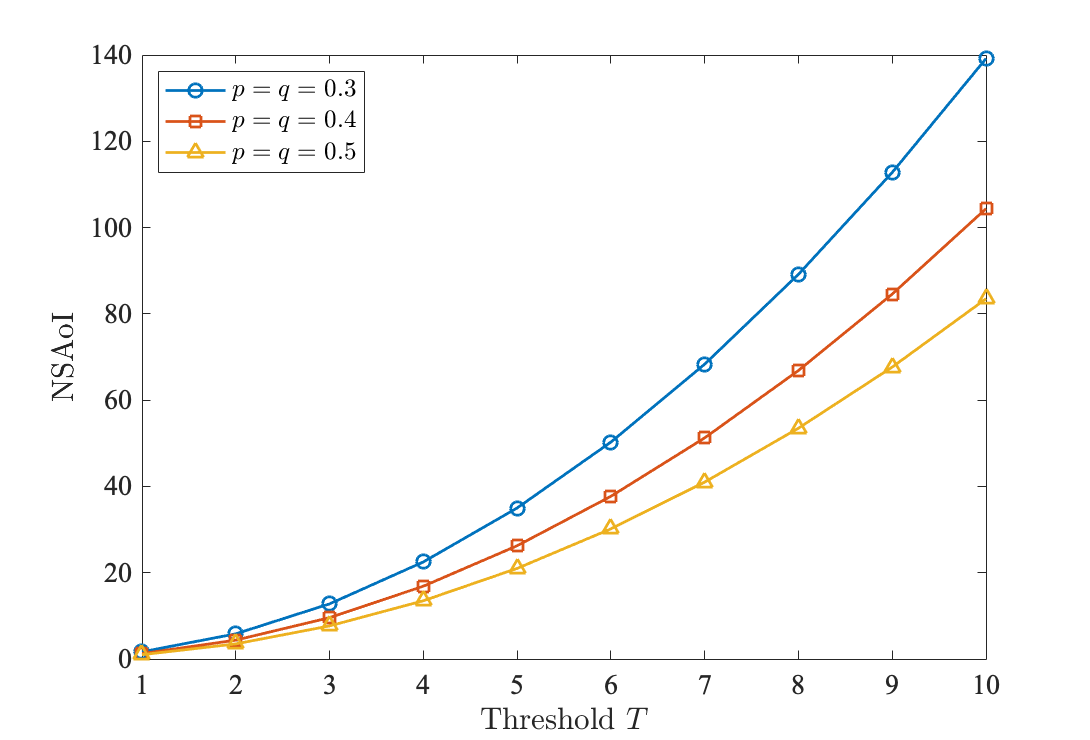}
    \caption{Normalized sum AoI versus the threshold $T$ for $p=q=0.3$, 0.4, 0.5.}
    \label{fig:aoi-T1}
\end{figure}

Fig.~\ref{fig:aoi-T1} shows the NSAoI as calculated using \eqref{NSAoI} for different sets of $(p,q)$. A larger $T$ results in a lower update rate as indicated by \eqref{update rate}, but
a higher NSAoI as longer update cycles have a higher average AoI per slot. 
Fig.~\ref{fig:aoi-T1} shows the NSAoI when $p=q$, i.e., a symmetric random walk. For a given $T$, the larger the $1-p-q$, the worse the NSAoI. The reason for this is that with a higher probability of staying in the same state for a symmetric random walk, it is more likely that the state remains the same, requiring a longer time to reach the boundaries.

\begin{figure}[!t]
\centering
    \includegraphics[scale=0.26]{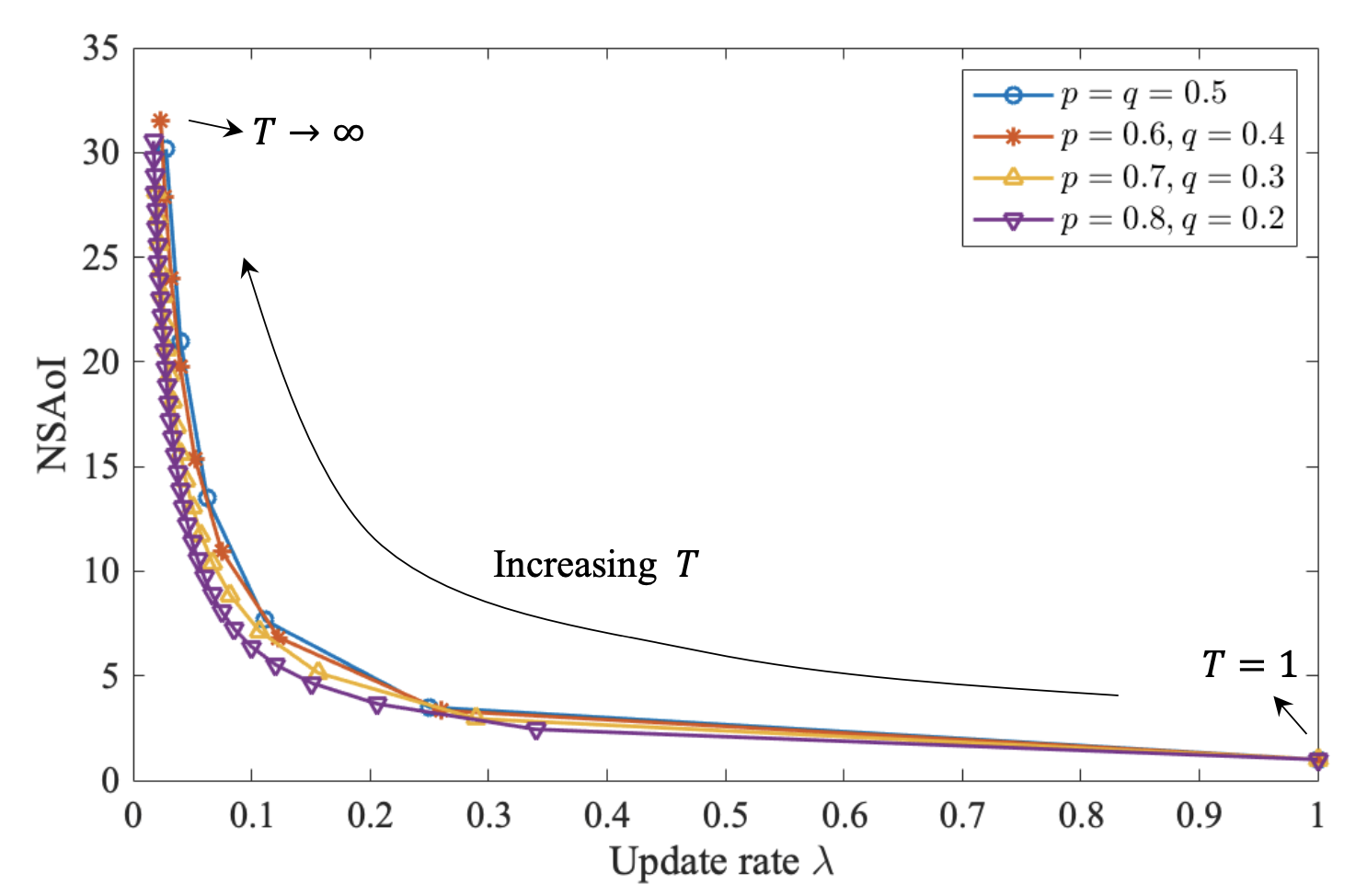}
    \caption{Normalized sum AoI versus update rate $\lambda$ for different sets of $(p,q)$ values.}
    \label{fig:first}
\end{figure}

The NSAoI versus update rate as $T$ varies is plotted in Fig.~\ref{fig:first}.
As $T\to\infty$, the update rate goes to 0, but the NSAoI increases without bound. In contrast, when $T=1$ (its minimum value), the update rate reaches its maximum value, and the NSAoI reaches its minimum value.

\subsection{Accuracy of Information}
\label{subsec:numerical-mse}
\begin{figure}[!t]
\centering
\includegraphics[scale=0.21]{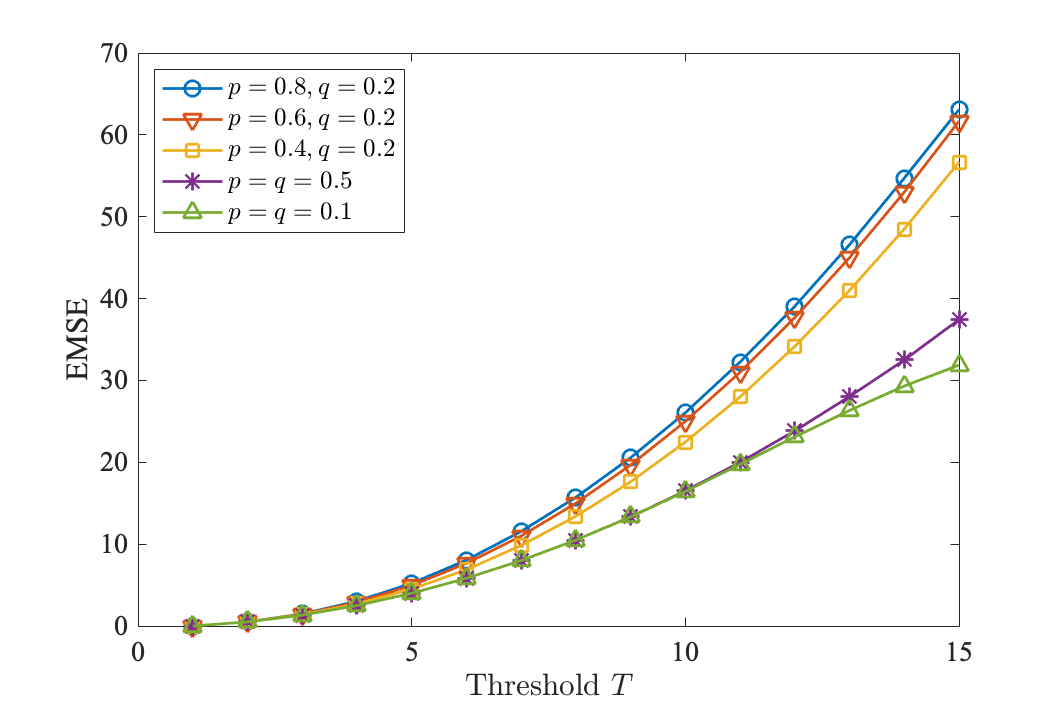}
\caption{Expected long-term MSE as a function of threshold $T$ for different sets of $(p,q)$ values.}
\label{fig:mse-plot}
\end{figure}
Fig.~\ref{fig:mse-plot} shows the long-term EMSE as calculated using \eqref{emse-cal}. As $T$ increases, the EMSE increases because a lower update rate results in a higher probability of drifting away from the initial state. When $T=1$, the EMSE is 0 for all sets of $(p,q)$ values. The reason is that, for $p+q=1$, in every time slot the state changes and an update is sent to the destination, which keeps the destination synchronized. For $p+q<1$, if the state changes, the destination is synchronized; if the state remains the same, the destination estimates the state to be the same as what it received last time, making the estimation equal to the true value. 
When $T \to \infty$, the EMSE also goes to infinity as the boundaries are never reached, leading to an unbounded increase in the error.
In addition, for a given $T$, the more asymmetric the random walk, the worse the EMSE. This is because, with a higher probability of drifting in one direction, it is more likely that the states in a shifted update cycle are  further away from the initial state in this cycle.
For a small $T$ ($T<10$), we observe that the EMSE is similar for symmetric random walks, i.e., $p=q$, $p+q \in [0,1]$. This is because the stationary distributions for $p=q$ are very similar when $T$ is small.

\subsection{Minimum update rate}
We have seen that both the NSAoI and the EMSE decrease with the update rate $\lambda$. 
Given maximum tolerable values for the NSAoI and the EMSE, we can determine the minimum required update rate. We illustrate this with an example.

\begin{figure}[!t]
\centering
\includegraphics[scale=0.25]{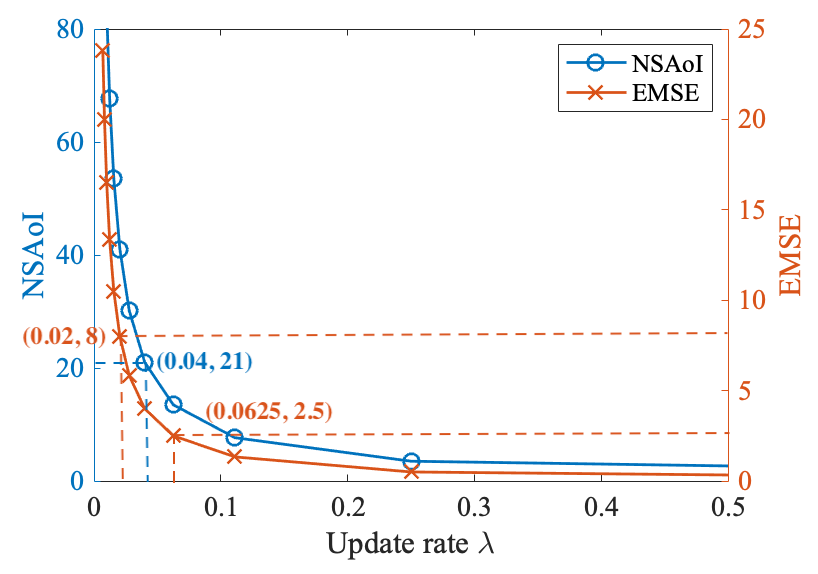}
\caption{NSAoI and EMSE versus the update rate, $\lambda$, for $p=q=0.5$.}
\label{fig:3d}
\end{figure}

Fig.~\ref{fig:3d} shows the three metrics, with the update rate $\lambda$ on the x-axis, the NSAoI on the left y-axis, the EMSE on the right y-axis. 
If an application requires the NSAoI to be no greater than 21 and the EMSE to be no greater than 2.5, we can 
obtain two corresponding minimum values of $\lambda$: 0.04 and 0.0625, as shown in Fig.~\ref{fig:3d}. Thus, to meet both requirements, the minimum required update rate should be chosen as 0.0625. 
Similarly, if an application requires the NSAoI to be no greater than 21 and the EMSE to be no greater than 8, the minimum required update rate is 0.04, as shown in Fig.~\ref{fig:3d}.

\section{Conclusions}
\label{sec:conclusions}
We studied a wireless network with a single sensor transmitting time-sensitive information to a destination. We modelled the sensor data as a 1-D random walk and used two boundaries defined by a threshold $T$ to adjust the update rate. The sensor sends an update to the destination whenever the random walk reaches either boundary, i.e. the state has changed by $\pm T$, since the last update.



We analyzed the system performance in terms of 3 metrics: update rate, normalized sum AoI (NSAoI), and expected mean squared error (EMSE). The analytic results which were verified using computer simulation can be used in the design of such systems. For example, they can be used to determine the minimum update rate required to meet specific requirements on NSAoI and EMSE.

\section*{ACKNOWLEDGMENTS}
This work was supported in part by the Natural Sciences and Engineering Research Council (NSERC) of Canada under Grants RGPIN-2020-05410 and RGPIN 1731-2013 and by the UBC PMC-Sierra Professorship in Networking and Communications. The authors would like to thank Dr I. Kadota for his advice on reproducing some of the simulation results in his paper \cite{kadota_1}.


\bibliographystyle{IEEEtran}
\bibliography{biblio}
\vspace{12pt}

\end{document}